# Local circular polarizations in nanostructures induced by linear polarization via optical near-fields

Makoto Naruse[1,*], Takeharu Tani[2], Tetsuya Inoue[3], Hideki Yasuda[2], Hirokazu Hori[4], and Masayuki Naya[2]

1 Photonic Network Research Institute, National Institute of Information and Communications Technology, 4-2-1 Nukui-kita, Koganei, Tokyo 184-8795, Japan

2 Frontier Core-Technology Laboratories, Research and Development Management Headquarters, Fujifilm Corporation, Nakanuma, Minamiashigara, Kanagawa 250-0193, Japan

3 Yamanashi Industrial Technology College, Enzan, Kosyu, Yamanashi 404-0042, Japan

4 Interdisciplinary Graduate School of Medical and Engineering, University of Yamanashi, Kofu 400-8511, Japan

a) Corresponding author: naruse@nict.go.jp

**Abstract:** We previously reported [Naruse, *et al.* Sci. Rep. 4, 6077, 2014] that the geometrical randomness of disk-shaped silver nanoparticles, which exhibit high reflection at near-infrared wavelengths, serves as the origin of a particle-dependent localization and hierarchical distribution of optical near-fields in the vicinity of the nanostructure. In this study, we show that the induced polarizations are circular, particularly at resonant wavelengths. We formulate optical near-field processes between nanostructures, accounting for their polarizations and geometries, and attribute circular polarization to the layout-dependent phase difference between the electrical susceptibilities associated with longitudinal and transverse-electric components. This study clarifies the fundamental optical properties of random nanostructured matter and offers generic theoretical concepts for implementing nanoscale polarizations of optical near-fields.

**OCIS codes:** (260.5430) Polarization, (160.3918) Metamaterials, (160.4236) Nanomaterials

# 1. INTRODUCTION

Many optical metamaterials often contain well-defined shape-engineered nanostructures;[1,2] however, randomly organized nanostructures also display interesting characteristics[3,4] and have been usefully employed in applications in industry.[5,6] For example, the large-area, mass-producible, thin-film device, called Nano Silver Pavement (NASIP), blocks heat transfer from sunlight. NASIP, which comprises randomly distributed disk-shaped silver nanoparticles, highly reflects near-infrared (NIR) light and has been marketed as an energy-conserving device for cooling rooms during the summer months.[5-7]

In our previous study,[7] we analyzed the structural randomness of the silver nanoparticles by electromagnetic calculations and a theory of optical near-fields based on an angular spectrum. The calculations revealed localized and hierarchical optical near-fields, which are not observed in uniformly arranged nanostructures. By calculating the imbalance of the horizontal and vertical electron charge distribution induced in each of the nanostructures, we determined the *effective dipole*, defined as $\boldsymbol{d}^{(i)} = d^{(i)} \exp(i\phi^{(i)})$.[7] Here, $i$ indexes the nanostructures and $d^{(i)}$ and $\phi^{(i)}$ ($-\pi < \phi^{(i)} \leq \pi$) respectively denote the magnitude and spatial phase of $\boldsymbol{d}^{(i)}$.

In this paper, we extend our analysis to show that the dipoles induced in the nanostructures are actually circularly polarized by linearly polarized light irradiation. To understand this phenomenon, we analytically formulate the optical near-field interactions between two nanostructures, accounting for their polarizations and geometrical features. We demonstrate that the electrical susceptibilities of the longitudinal (L) and transverse-electric (TE) components exhibit a layout-dependent phase difference that leads to local circular polarizations. Ohdaira *et al.* successfully generated local circular polarizations by superposing two cross-propagating evanescent waves.[8] Local circular polarization would also enable the manipulation of nanometer-sized electronic systems via spin-orbit interactions[9] and the control of polarization-preferred chemical reactions. The presented theoretical background on shape-engineered nanostructure approach can help achieve the generation of local circular polarizations. Moreover, as demonstrated below, our theoretical treatment is fairly generalized, and therefore is applicable to other problems dependent on nanometer-scale polarizations and geometries. Notably, the dipole–dipole interactions between nanoparticles have been well-studied in conventional near-field optics,[10-12] and they have been exploited in optical near-field microscopy, as well as other novel functions such as hierarchical information retrieval.[13] Nevertheless, to our knowledge, circular polarizations generated by optical near-field interactions have not been theoretically formulated.

The remainder of this paper is organized as follows. Section 2 presents a detailed numerical analysis of the polarizations induced in random silver nanostructures. Section 3 develops an analytical theory of optical near-field interactions between two nanostructures and discusses the circular polarizations induced by linearly polarized light irradiation. Section 4 concludes the paper.

## 2. ELECTROMAGNETIC ANALYSIS OF THE NASIP DEVICE

### A. Review of the NASIP device and former analysis

First, we briefly review the silver nanoparticle-based NASIP device[5,6] and our former numerical analysis that highlighted the inherent structural randomness of this device.[7] Figure 1(a) shows a scanning electron microscope (SEM) image of the NASIP surface. The device exhibits high reflectance in the NIR regime while strongly transmitting visible and far-infrared light. Therefore, its translucence is sufficient, and it does not inhibit wireless communications. The elemental nanostructure is 120–150 nm in diameter and 10 nm thick and resonates at NIR wavelengths (centered around 1000 nm).[5-7]

To characterize the detailed electromagnetic properties of the experimentally fabricated devices, we input the geometries of the fabricated silver nanoparticles to a numerical model comprising a vast number of voxels. Specifically, the SEM image shown in Fig. 1(a), which occupies an area of 4.2 μm × 4.2 μm, is digitized into binary values with a horizontal and vertical resolution of 2.5 nm. Pixels occupied by silver nanoparticles and substrate material are assigned values of one and zero, respectively. Subsequently, the silver nanoparticles (particle number $N$ = 468), are numerically modeled in an $x$–$y$–$z$ Cartesian coordinate system containing 4200 × 4200 × 5 voxels, or 88.2 M voxels. This model is executed by a finite-difference time-domain-based electromagnetic simulator, assuming continuous-wave $x$-polarized light normally incident on the surface of the silver nanostructures. Figure 1(b) shows the electromagnetic intensity distribution at a distance of 5 nm from the surface of the silver nanoparticles. The surface is irradiated with 1000-nm incident light.

Formerly, by evaluating the statistical properties of the dipoles induced in the nanostructures, we characterized the structural randomness of the NASIP and its impact on the associated localized optical near-fields.[7] Specifically, we derived the induced charge distributions $\rho(x, y)$ by calculating the divergences of the electric fields within the silver nanostructures and by summing the distributions along the $z$ direction. Then, we derived the *effective dipole* $\boldsymbol{d}^{(i)}$ induced in each nanoparticle as the imbalance of electron charge with respect to the geometrical center of gravity:

$$\boldsymbol{d}^{(i)} = \left( \sum_{x \geq G_x^{(i)}} \rho(x,y) - \sum_{x < G_x^{(i)}} \rho(x,y), \sum_{y \geq G_y^{(i)}} \rho(x,y) - \sum_{y < G_y^{(i)}} \rho(x,y) \right) = d^{(i)} \exp(i\phi^{(i)}), \qquad (1)$$

where $\left( G_x^{(i)}, G_y^{(i)} \right)$ denotes the center of gravity of nanoparticle $i$. The statistics of the magnitude $d^{(i)}$ and (space-domain) phase $\phi^{(i)}$ were analyzed from 300 nm to 2000 nm at 100 nm intervals and were theoretically investigated using the angular spectrum representation of the optical near-fields.[7]

**B. Detailed electromagnetic analysis**

In our former study, we quantified the fundamental characteristics of the localized and hierarchical optical fields originating from structural randomness. Because we neglected the time-domain phase differences between the $x$ and $y$ components of the electric fields, as shown by Eq. (1), all our calculated effective dipoles were linear polarizations. To reveal whether local circular polarizations can be generated, we revised the analysis as follows:

The induced dipoles are now calculated by:

$$\begin{aligned} \boldsymbol{q}^{(i)} &= \int \boldsymbol{r}\rho(\boldsymbol{r})d\boldsymbol{r} \\ &= (|q_x|\exp[i\arg(q_x)], |q_y|\exp[i\arg(q_y)]), \end{aligned} \quad (2)$$

where the time-domain phase difference $\phi = \arg(q_y) - \arg(q_x)$ can be non-zero. The integral in Eq. (2) is computed over the area occupied by particle $i$. Note that there is no need to subtract a "center of gravity of the charge distribution" in Eq. (2) because the total induced charge is zero. Physically, Eq. (2) describes a dipole oscillating through an ellipse with long and short axes $B_1$ and $B_2$, respectively. The $B_1$ axis is tilted by angle $\psi$ to the $x$-axis, as schematized in Fig. 2(a).[14] The degree of circular polarization is estimated by the ellipticity:

$$\frac{B_2}{B_1} = \tan\chi. \quad (3)$$

We also denote counterclockwise rotation by $\sin(\phi) > 0$. A movie of the electron charge distributions within a unit interval of light oscillation (wavelength = 1000 nm) is provided in the supplementary material. The circularly oscillating dynamics of the electron charges are clearly observed in this movie. For deriving the ellipticity given by Eq. (3) from the quantities in Eq. (2), the reader is referred to Born and Wolf.[14]

Figure 2(b) is a schematic of the induced dipoles calculated under 1000 nm incident light. Blue and red ellipses rotate counterclockwise and clockwise, respectively, and their sizes indicate their amplitudes ($B_1$ and $B_2$). Circular (strictly elliptic) polarizations are clearly observed in this figure. From the histogram of the ellipticities (Fig. 2(c)), we find that most of the dipoles have small ellipticity; however, some of them are highly elliptic (with ellipticities of plus or minus unity). The average and standard deviation of the ellipticity are separately plotted as functions of wavelength in Fig. 2(d). The standard deviation increases around the resonant wavelengths.

We consider that such circular polarizations emerge from structural randomness and resonance between the dipoles and the input light. In the next section, we present an analytical formulation of these phenomena.

## 3. THEORY FOR CIRCULAR POLARIZATION VIA OPTICAL NEAR-FIELDS

### A. Theory of near-field interactions

Our theoretical analysis assumes two metal nanoparticles ($S_1$ and $S_2$) of radius $a$. The centers of $S_1$ and $S_2$ are placed at the origin of the Cartesian system and at $\boldsymbol{R}_{21}$, respectively, as shown in Fig. 3. Recall that disk-shaped nanoparticles are assumed in the experimental NASIP device and the analysis of Section 2; therefore, nanoparticle-based modeling does not exactly match with the NASIP device architecture. Nevertheless, we consider that the nanoparticle-based approach of our study preserves the essential attributes of the system and reveals the essential underling physical processes.

When the size and constituents of two particles are identical, the polarizations induced in $S_1$ and $S_2$ are also identical and are given by:

$$\boldsymbol{P}_t = \left(\frac{6\pi\varepsilon_0}{iK^3}\right) \sum_{\mu=L,TM}^{TE} \left[\frac{\alpha_1^{(E)}}{1-G_\mu(KR_{21})\alpha_1^{(E)}}\right] \boldsymbol{\varepsilon}_\mu(\boldsymbol{n}_{21})[\boldsymbol{\varepsilon}_\mu(\boldsymbol{n}_{21}) \bullet \boldsymbol{E}^{(0)}(0)]. \tag{4}$$

When the two nanoparticles are in a subwavelength regime, the interactions are called optical near-field interactions. In Eq. (4), the Mie scattering coefficients of $S_1$ and $S_2$ are equal and are given by $\alpha_1^{(E)}$. The general form and related details of Eq. (4) are provided in Appendix A. The unit polarizations vectors $\varepsilon_\mu(\boldsymbol{n}_{21})$ are expressed in polar coordinates as

$$\begin{aligned}\varepsilon_L(\boldsymbol{n}_{21}) &= \boldsymbol{n}_{21} = (\sin\theta\cos\phi, \sin\theta\sin\phi, \cos\theta) \\ \varepsilon_{TM}(\boldsymbol{n}_{21}) &= (\cos\theta\cos\phi, \cos\theta\sin\phi, -\sin\theta) \quad , \\ \varepsilon_{TE}(\boldsymbol{n}_{21}) &= (-\sin\phi, \cos\phi, 0)\end{aligned} \tag{5}$$

where $\boldsymbol{n}_{21}$ is the unit vector $\boldsymbol{R}_{21}/R_{21}$, $\boldsymbol{E}^{(0)}(0)$ is the complex amplitude of the input light generated at the center of $S_1$, and $G_\mu(KR_{21})$ is the electric dipole propagator from $S_1$ to $S_2$, calculated by Eq. (6) below (the general form is derived in Appendix A). When the distance between $S_1$ and $S_2$ is much smaller than the light wavelength ($KR_{21} \ll 1$), the electric dipole propagators in Eq. (4) are given by:

$$\begin{aligned}G_L(KR_{21}) &\sim -3i\frac{1}{(KR_{21})^3} \\ G_{TM}(KR_{21}) = G_{TE}(KR_{21}) &\sim -\frac{1}{2}G_L(KR_{21})\end{aligned} \tag{6}$$

and the Mie scattering coefficients are approximated by:

$$\alpha_1^{(E)} \sim \frac{2}{3}i\left(\frac{n^2-1}{n^2+2}\right)(Ka)^3. \tag{7}$$

Substituting Eqs. (6) and (7) in (4), we obtain:

$$\boldsymbol{P}_t = \varepsilon_0 \Delta V \sum_{\mu=L,TM}^{TE} \chi_\mu(KR_{21})\varepsilon_\mu(\boldsymbol{n}_{21})[\varepsilon_\mu(\boldsymbol{n}_{21}) \bullet \boldsymbol{E}^{(0)}(0)], \tag{8}$$

where $\Delta V$ is the nanoparticle volume. Equation (8) can be interpreted as the sum of polarizations with different electrical susceptibilities given by:

$$\chi_\mu(KR_{21}) = B_\mu \left( \frac{n^2 - 1}{n^2 + A_\mu} \right), \tag{9}$$

where

$$A_\mu = \frac{2 - \sigma_\mu}{1 + \sigma_\mu}, \quad B_\mu = \frac{3}{1 + \sigma_\mu}. \tag{10}$$

Here, $\sigma_\mu$ is the ratio of the nanoparticle radius and the separation between $S_1$ and $S_2$:

$$\sigma_L = -2\left(\frac{a}{R_{21}}\right)^3, \quad \sigma_{TM} = \sigma_{TE} = -\frac{1}{2}\sigma_L. \tag{11}$$

With a separation of the real and imaginary parts of the refractive index given by $n = \tilde{n} + i\tilde{k}$, the electrical susceptibility given in Eq. (9) is expressed as

$$\chi_\mu(KR_{21}) = B_\mu \left\{ 1 - (1 + A_\mu) \left[ \frac{(\tilde{n}^2 - \tilde{k}^2 + A_\mu)}{(\tilde{n}^2 - \tilde{k}^2 + A_\mu)^2 + (2\tilde{n}\tilde{k})^2} \right] \right. \\ \left. + i(1 + A_\mu) \left[ \frac{(2\tilde{n}\tilde{k})}{(\tilde{n}^2 - \tilde{k}^2 + A_\mu)^2 + (2\tilde{n}\tilde{k})^2} \right] \right\}. \tag{12}$$

From Eq. (12), we find that the electrical susceptibility in the L basis differs from that in the transverse magnetic/TE basis.

If $S_1$ and $S_2$ are located on the $x$–$y$ plane, the orientation of $S_2$ with respect to $S_1$ is given by $(\theta, \phi) = (\pi/2, \phi)$ (see Eq. (5)). In this situation, Eq. (8) implies that an $x$-polarized input light will couple with the L and TE bases of the optical near-field interactions. The polarization in Cartesian coordinates is given by $\boldsymbol{P}_t = P_x \boldsymbol{e}_x + P_y \boldsymbol{e}_y$, where:

$$\begin{aligned} P_x &= \varepsilon_0 \Delta V E^{(0)} \left[ \chi_+(KR_{21}) + \chi_-(KR_{21}) \cos 2\phi \right] \\ P_y &= \varepsilon_0 \Delta V E^{(0)} \chi_-(KR_{21}) \sin 2\phi \end{aligned} \tag{13}$$

with

$$\chi_{\pm}(KR_{21}) = \frac{1}{2}[\chi_L(KR_{21}) \pm \chi_{TE}(KR_{21})]. \tag{14}$$

**B. Local circular polarization**

We first characterize the polarizations induced by *x*-polarized light by examining some representative cases using the theory developed in Section 3.A.

(i) When $S_1$ and $S_2$ are horizontally aligned (i.e., $\phi = 0$ ), $P_y$ in Eq. (13) vanishes because the input light couples only with the L basis (implying no circular polarization).

(ii) When $S_1$ and $S_2$ are vertically aligned, (i.e., $\phi = \pi/2$ ), $P_y$ vanishes because the input light couples only with the TE basis (again implying no circular polarization).

Along with the simple examples (i) and (ii) above, Eq. (13) indicates that whenever $P_y$ is non-zero, the input light must be coupled to *both* L and TE polarization bases, generating different electrical susceptibilities $\chi_L$ and $\chi_{TE}$.

By explicitly expressing the complex number $P_j$ ( $j = x, y$) in Eq. (13) as $P_j = |P_j| \exp(i\delta_j)$, we classify the trajectory of oscillations as circular or linear by the following metrics:

$$\xi = \frac{|P_y|}{|P_x|} \tag{15}$$

$$\cos\eta = \cos(\delta_x - \delta_y). \tag{16}$$

[A] When $|\cos\eta| < 1$, the trajectory is elliptic (or completely circular when $\cos\eta = 0$ and $\xi = 1$), namely, circular polarization is induced.

[B] When $\cos\eta = \pm 1$, the trajectory is a line segment; i.e., the induced polarization is linear.

Cases (i) and (ii) above are examples of linear polarization (category [B]). In addition, if the nanoparticles constitute a dielectric material, both $\chi_L$ and $\chi_{TE}$ are real. In this case, the time-domain phase difference is zero; consequently, $\eta = 0$ and the polarization is again linear. Another category [B] case is two distantly separated nanoparticles with negligible optical near-field interactions. In this case, $\chi_L$ and $\chi_{TE}$ are equal and $y$-component is not present in the induced polarization (see Eq. (13)).

As a representative case of category [A], we investigate the polarization when $S_1$ and $S_2$ are aligned at $\phi = \pi/4$. According to Eq. (13), $P_x$ and $P_y$ depend only on $\chi_+$ and $\chi_-$, respectively. Hence, the phase difference between $P_x$ and $P_y$ corresponds to the difference between the electrical susceptibilities $\chi_+$ and $\chi_-$. Panels (a) and (b) of Fig. 4 plot the cosine of the phase difference (Eq. (15)) and the absolute value ratio of the $x$- and $y$-amplitude (Eq. (14)), respectively, as functions of the imaginary part of the refractive index ($\tilde{k}$). Here, the real part of the refractive index is fixed at $\tilde{n} = 0.2$. The ratio of the particle radius to the inter-particle distance ($a/R_{21}$) is varied as 1/3, 1/4, and 1/5. As shown in Fig. 4(a), $|\cos\eta| < 1$ is satisfied in the parameter regime around the plasmon resonance; i.e., $\tilde{k} \sim \tilde{k}_p = \sqrt{\tilde{n}^2 + 2}$ and the polarization is circular. From Fig. 4(b), we can find that the circular polarization is elongated in the horizontal direction ($\xi < 0.5$). As the inter-particle distance increases, the ellipticity deteriorates and the polarization eventually becomes linear. The curves in Fig. 4(c) are the oscillating trajectories when $\phi$ is varied as $\pi/16$, $\pi/8$, $\pi/4$, $3\pi/8$, and $7\pi/16$. The degree of circular polarization is maximized at $\phi = \pi/4$, as expected because $P_y$ is proportional to $\sin 2\phi$ (see Eq. (13)).

As described earlier, the phase difference between $P_x$ and $P_y$ corresponds to the difference between $\chi_+$ and $\chi_-$ when $\phi = \pi/4$. The four curves in Fig. 4(d) plot the real and imaginary parts of $\chi_+$ and $\chi_-$ as functions of the imaginary part of the refractive index, fixing $\tilde{n} = 0.2$ and $a/R_{21} = 1/3$. From Fig. 4(d), we recognize three regimes of the emerging phase difference:

1. In the regime $\tilde{k} \sim \tilde{k}_p$,

   Re[$\chi_+$] ~ 0, Im[$\chi_+$] > 0, Re[$\chi_-$] < 0, and Im[$\chi_-$] ~ 0.

   In this regime, the phase difference between $\chi_+$ and $\chi_-$ is $-\pi/2$, and thus $\cos\eta = 0$.

2. In the regime $\tilde{k} > \tilde{k}_p$,

   Re[$\chi_+$] > 0, Im[$\chi_+$] ~ 0, Re[$\chi_-$] is small and positive, Im[$\chi_-$] ~ 0. In this regime, the phase difference between $\chi_+$ and $\chi_-$ is 0, and thus $\cos\eta = 1$.

3. In the regime $\tilde{k} \sim \tilde{k}_p$,

   Re[$\chi_+$] < 0, Im[$\chi_+$] ~ 0, Re[$\chi_-$] is small and positive, Im[$\chi_-$] ~ 0. In this regime, the phase difference between $\chi_+$ and $\chi_-$ is $\pi$, and thus $\cos\eta = -1$.

From these findings, we conclude that linearly polarized input light can induce circular polarization in nanostructures when it couples with both the L and TE polarization bases in regions of different electrical susceptibility at frequencies around the plasmon resonances.

We suggest one minor remark regarding the definition of ellipticity in the numerical analysis in Section 2 and theoretical analysis in Section 3. The $B_2/B_1$ in Eq. (3) and $\xi = |P_y|/|P_x|$ in Eq. (15) are physically similar to each other, and $\phi$ associated with Eq. (2) and $(\delta_x - \delta_y)$ in Eq. (16) are also similar. However, these are *not* necessarily equal to each other. As shown by the case [B], $\xi = |P_y|/|P_x|$ can be unity while the phase difference $(\delta_x - \delta_y)$ is zero, which is a linearly polarized light. On the other hand, such a linear polarization is expressed as $B_2/B_1 = 0$ in the definition in Section 2.

### C. Local circular polarization in three particle systems

When the number of nanoparticles ($N$) exceeds three, the theoretical analysis becomes intractable. This subsection analyses the case of $N = 3$, assuming that the total polarization induced at a particular particle is the sum of the individual interactions between the particle and the other two particles. This assumption approximates the real situation.

In the system of Fig. 5(a), $S_1$ is located at the Cartesian origin, and $S_2$ and $S_3$ are oriented at $\phi = \pi/4$ and $-\pi/4$ with radius-to-distance ratios of $a/R_{21} = a/R_{31} = 1/3$, respectively. This configuration is referred to as a "symmetric" configuration. The polarization induced in $S_1$ is derived by considering the interactions between [$S_1$ and $S_2$] and [$S_1$ and $S_3$]. We find that although circular polarizations are induced in $S_2$ and $S_3$, $S_1$ is linearly polarized (Fig. 5(b)). This outcome is attributed to the *symmetric* layout of the nanoparticles with respect to the *x*-polarization, which cancels the circular polarizations at $S_1$. The system shown in Fig. 5(c), on the other hand, is asymmetric about the *x* axis. In this layout, $S_2$ and $S_3$ are oriented by $\phi = \pi/4$ and $\phi = -\pi/8$, respectively, with respect to $S_1$, which is referred to as an "asymmetric" configuration. As depicted in Fig. 5(d), $S_1$, $S_2$, and $S_3$ are circularly polarized. These results suggest that circular polarizations are induced in *asymmetric* nanoparticle layouts. Such mechanisms may explain the experimentally observed circular polarizations discussed in Section 2.

To support our analyses, we performed additional finite-difference time-domain simulations to those described in Section 2, assuming similar conditions. Specifically, we arranged silver disk-shaped nanoparticles (of diameter and thickness 120 nm and 10 nm, respectively) into symmetric and asymmetric configurations. The inter-disk distance was 50 nm, and the layouts were subjected to 1000-nm *x*-polarized light. Figure 6(a) and (b) shows the optical intensity distributions in symmetric and asymmetric configurations, respectively, evaluated on the plane equidistant from both silver nanostructure surfaces. Following the analysis method discussed in Section 2, we evaluated the absolute value of the ellipticity (given by Eq. (3)) in particles $S_1$, $S_2$, and $S_3$. The results, plotted in Fig. 6(c), well-agree with theory, which predicts near-zero ellipticity of $S_1$ in the symmetric configuration but only elliptical polarizations in the asymmetric configuration. Precisely, the ellipticity of $S_1$ in the symmetric configuration is in the order of $10^{-5}$, not completely zero. We consider this as a numerical artifact mainly because of the fact that the numerical modeling assumes a disk-shaped architecture.

## 4. CONCLUSION

In this paper, we showed that polarizations induced in disk-shaped randomly organized silver nanoparticles that highly reflect at NIR wavelengths are circularly polarized, particularly at resonant wavelengths. We theoretically examined such phenomena by considering the optical near-field processes between nanostructures, accounting for their polarizations and geometries. Depending on the nanostructure layout, a phase difference emerges between the electrical susceptibilities associated with the L and TE polarization components, which appears to source the circular polarizations. Light input to asymmetric (or random) nanoparticle arrangements generates circular polarization in each of the elemental nanostructures in the system. This study provides a fundamental insight into the optical properties of random nanostructured matter while offering generic theoretical concepts for implementing nanoscale polarizations of optical near-fields.

**Acknowledgements**

The authors thank M. Ohtsu for fruitful discussions. This work was supported in part by the Core-to-Core Program A. Advanced Research Networks from the Japan Society for the Promotion of Science.

**References**

1. M. Kadic, T. Bückmann, R. Schittny, and M. Wegener, "Metamaterials beyond electromagnetism," Rep. Prog. Phys. **76**, 126501 (2013).

2. J. Valentine, S. Zhang, T. Zentgraf, E. Ulin-Avila, D. A. Genov, G. Bartal, and X. Zhang, "Three-dimensional optical metamaterial with a negative refractive index," Nature **455**, 376-379 (2008).

3. B. Redding, M. A. Choma, and H. Cao, "Speckle-free laser imaging using random laser illumination," Nat. Photonics **6**, 355-359 (2012).

4. S. Grésillon, L. Aigouy, A. C. Boccara, J. C. Rivoal, X. Quelin, C. Desmarest, P. Gadenne, V. A. Shubin, A. K. Sarychev, and V. M. Shalaev, "Experimental observation of localized optical excitations in random metal-dielectric films," Phys. Rev. Lett. **82**, 4520-4523 (1999).

5. N. Kiyoto, S. Hakuta, T. Tani, M. Naya, and K. Kamada, "Development of a Near-infrared Reflective Film Using Disk-shaped Silver Nanoparticles," Fujifilm Res. and Dev. **58-2013**, 55-58 (2013).

6. T. Tani, S. Hakuta, N. Kiyoto, and M. Naya, "Transparent near-infrared reflector metasurface with randomly dispersed silver nanodisks," Opt. Express **22**, 9262- 9270 (2014).

7. M. Naruse, T. Tani, H. Yasuda, N. Tate, M. Ohtsu, and M. Naya, "Randomness in highly reflective silver nanoparticles and their localized optical fields," Sci. Rep. **4**, 6077 (2014).

8. Y. Ohdaira, T. Inoue, H. Hori, and K. Kitahara, "Local circular polarization observed in surface vortices of optical near-fields," Opt. Express **16**, 2915-2921 (2008).

9. S. F. Alvarado and P. Renaud, "Observation of spin-polarized-electron tunneling from a ferromagnet into GaAs," Phys. Rev. Lett. **68**, 1387-1390 (1992).

10. M. Ohtsu and H. Hori, *Near-Field Nano-Optics* (Kluwer/Plenum, 1999).

11. H. Tamaru, H. Kuwata, H. T. Miyazaki, and K. Miyano, "Resonant light scattering from individual Ag nanoparticles and particle pairs," Appl. Phys. Lett. **80**, 1826-1828 (2002).

12. H. Xu, H., E. J. Bjerneld, M. Käll, and L. Börjesson, "Spectroscopy of Single Hemoglobin Molecules by Surface Enhanced Raman Scattering," Phys. Rev. Lett. **83**, 4357-4360 (1999).

13. M. Naruse, T. Yatsui, W. Nomura, N. Hirose, and M. Ohtsu, "Hierarchy in optical near-fields and its application to memory retrieval," Opt. Express **13**, 9265-9271 (2005).

14. M. Born and E. Wolf, *Principles of Optics* (Cambridge, 1999).

**Figure captions**

**Fig. 1.** (a) Scanning electron microscopy (SEM) image of the fabricated near-infrared light reflection film composed of silver nanoparticles (NASIP (Nano Silver Pavement)). (b) Electric field intensity distributions, calculated at the distance of 5 nm from the silver nanostructure. The wavelength of the normally incident light is 1000 nm.

**Fig. 2.** (a) Schematic of ellipsoidal polarization. The degree of circular polarization is evaluated by the ellipticity $B_2 / B_1$. (b) Calculated distributions of the polarizations induced in each of the nanostructures under 1000-nm incident light. The blue and red circles respectively denote counterclockwise and clockwise rotations. (c) Histogram of the ellipticities in (b). (d) Statistical properties of the ellipticity as functions of wavelength. The standard deviation increases at resonant wavelengths (~1000 nm).

**Fig. 3.** Schematic of the Cartesian coordinate system adopted in the theory. Spherical nanoparticle $S_2$ is displaced from $S_1$ (located at the origin) by $\bm{R}_{21}$. The angle $\phi$ is the angle between $\bm{R}_{21}$ and the $x$-axis when $S_1$ and $S_2$ lie on the $x$–$y$ plane.

**Fig. 4.** (a) Cosine of the phase difference between $P_x$ and $P_y$ and (b) absolute value ratio of $P_x$ to $P_y$, as functions of the imaginary part of the refractive index. The real part of the refractive index is 0.2. The solid, dashed, and dotted curves denote relative inter-particle separations ($a / R_{21}$) of 1/3, 1/4, and 1/5, respectively. (c) Polarization trajectories for different orientations of the two particles, maintaining the inter-particle distance $a/R_{21} = 1/3$. (d) Real and imaginary parts of $\chi_+$ and $\chi_-$ as functions of the imaginary part of the refractive index.

**Fig. 5.** Three-particle systems and the polarizations induced in each particle. In (a) "Layout 1" and (b) "Layout 2," $S_2$ and $S_3$ are symmetrically and asymmetrically positioned either side of the *x*-axis, respectively. The distances between [$S_1$ and $S_2$] and [$S_1$ and $S_3$] are three times the particle radius. (c) and (d) summarize the theoretically-calculated induced polarizations under *x*-polarized input light. Particle $S_1$ is linearly polarized in Layout 1, whereas all particles are circularly polarized in Layout 2.

**Fig. 6.** (a, b) Electromagnetic simulations of three disk-shaped silver nanoparticles arranged in (a) symmetric and (b) asymmetric configurations. (c) Calculated ellipticity in both configurations. Square, circular, and triangular symbols denote the absolute values of the ellipticity in particles $S_1$, $S_2$, and $S_3$, respectively. As theoretically predicted, the ellipticity in $S_1$ is nearly zero in the symmetric configuration, whereas all ellipticities are non-zero in the asymmetric configuration.

**Appendix A**

The analysis of Section 3.A considered multiple transfers of polarizations between two nanoparticles. Let $\boldsymbol{P}^{(1)}$ and $\boldsymbol{P}^{(2)}$ be the dipoles induced by light irradiation on S$_1$ and S$_2$, respectively given by $\boldsymbol{P}^{(1)} = \left(6\pi\varepsilon_0/iK^3\right)\alpha_1^{(E)}(1)\boldsymbol{E}^{(0)}(\boldsymbol{0})$ and $\boldsymbol{P}^{(2)} = \left(6\pi\varepsilon_0/iK^3\right)\alpha_1^{(E)}(2)\boldsymbol{E}^{(0)}(\boldsymbol{R}_{21})$. Here, $\alpha_1^{(E)}(j)$ is the Mie scattering coefficient of particle S$_j$ ($j = 1, 2$). For example, the generated $\boldsymbol{P}^{(1)}$ excites another polarization in S$_2$, $\boldsymbol{P}'^{(2)}$, given by

$$\boldsymbol{P}'^{(2)} = \sqrt{\frac{3}{8\pi}} \alpha_1^{(E)}(2) \sum_{m=-1}^{1} \boldsymbol{U}_{K,1,m}^{(E)RA}(\boldsymbol{R}_{21}) \left[ \boldsymbol{e}_m^* \cdot \boldsymbol{P}^{(1)} \right], \tag{A.1}$$

where * means a complex conjugate. Here, $\boldsymbol{e}_m$ is described on a spherical basis given by $\boldsymbol{e}_0 = \boldsymbol{e}_z$, $\boldsymbol{e}_{\pm 1} = \mp 1/\sqrt{2}(\boldsymbol{e}_x \pm i\boldsymbol{e}_y)$.

In (A.1), the vector spherical wave is given by

$$\boldsymbol{U}_{K,1,m}^{(E)RA}(\boldsymbol{R}_{21}) = \sqrt{6\pi} \left\{ -\frac{1}{3} h_2^{(1)}(KR_{21})[\boldsymbol{e}_m - 3\boldsymbol{n}_{21}(\boldsymbol{n}_{21} \cdot \boldsymbol{e}_m)] + \frac{2}{3} h_0^{(1)}(KR_{21})\boldsymbol{e}_m \right\}, \tag{A.2}$$

where $h_0^{(1)}(KR_{21})$ and $h_2^{(1)}(KR_{21})$ are the zero- and second-order spherical Hankel functions, respectively. Substituting Eq. (A.2) in (A.1) and converting from spherical to polar coordinates (see Eq. (5)), $\boldsymbol{P}'^{(2)}$ is given by

$$\boldsymbol{P}'^{(2)} = \sum_{\mu=L,TM}^{TE} \boldsymbol{\varepsilon}_\mu(\boldsymbol{n}_{21}) \left[ \alpha_1^{(E)}(2) G_\mu(KR_{21}) \right] \left[ \boldsymbol{\varepsilon}_\mu(\boldsymbol{n}_{21}) \cdot \boldsymbol{P}^{(1)} \right], \tag{A.3}$$

where

$$\begin{aligned} G_L(KR_{21}) &= h_2^{(1)}(KR_{21}) + h_0^{(1)}(KR_{21}) \\ G_{TM}(KR_{21}) &= -\frac{1}{2}\left[ h_2^{(1)}(KR_{21}) - 2h_0^{(1)}(KR_{21}) \right]. \\ G_{TE}(KR_{21}) &= G_{TM}(KR_{21}). \end{aligned} \tag{A.4}$$

Likewise, $\boldsymbol{P}^{(2)}$ induces a polarization $\boldsymbol{P}'^{(1)}$ in $S_1$. Polarization $\boldsymbol{P}'^{(1)}$ then induces another polarization $\boldsymbol{P}''^{(2)}$ in $S_2$. Thus, the total polarization induced in $S_1$ and $S_2$ is given by

$$\boldsymbol{P}_t^{(j)} = \boldsymbol{P}^{(j)} + \boldsymbol{P}'^{(j)} + \boldsymbol{P}''^{(j)} + \boldsymbol{P}'''^{(j)} + \cdots (j=1,2). \tag{A.5}$$

Substituting Eq. (A.3) in (A.5), the total polarizations of $S_1$ and $S_2$ are respectively given by

$$\boldsymbol{P}_t^{(1)} = \sum_{\mu=L,TM}^{TE} \boldsymbol{\varepsilon}_\mu(\boldsymbol{n}_{21}) \left[ \frac{1}{1-Q_\mu(KR_{21})} \right] \left\{ \left[ \boldsymbol{\varepsilon}_\mu(\boldsymbol{n}_{21}) \bullet \boldsymbol{P}^{(1)} \right] + \alpha_1^{(E)}(1) G_\mu(KR_{21}) \left[ \boldsymbol{\varepsilon}_\mu(\boldsymbol{n}_{21}) \bullet \boldsymbol{P}^{(2)} \right] \right\} \tag{A.6}$$

and

$$\boldsymbol{P}_t^{(2)} = \sum_{\mu=L,TM}^{TE} \boldsymbol{\varepsilon}_\mu(\boldsymbol{n}_{21}) \left[ \frac{1}{1-Q_\mu(KR_{21})} \right] \left\{ \left[ \boldsymbol{\varepsilon}_\mu(\boldsymbol{n}_{21}) \bullet \boldsymbol{P}^{(2)} \right] + \alpha_1^{(E)}(2) G_\mu(KR_{21}) \left[ \boldsymbol{\varepsilon}_\mu(\boldsymbol{n}_{21}) \bullet \boldsymbol{P}^{(1)} \right] \right\}, \tag{A.7}$$

where

$$Q_\mu(KR_{21}) = \left[ \alpha_1^{(E)}(1) G_\mu(KR_{21}) \right] \left[ \alpha_1^{(E)}(2) G_\mu(KR_{21}) \right]. \tag{A.8}$$

When the radii and the numerical indexes of $S_1$ and $S_2$ are equal and the coefficients are given by $\alpha_1^{(E)}(1) = \alpha_1^{(E)}(2) = \alpha_1^{(E)}$, we obtain $\boldsymbol{P}_t^{(1)} = \boldsymbol{P}_t^{(2)} = \boldsymbol{P}_t$, and Eqs. (A. 6) and (A. 7) become Eq. (4).

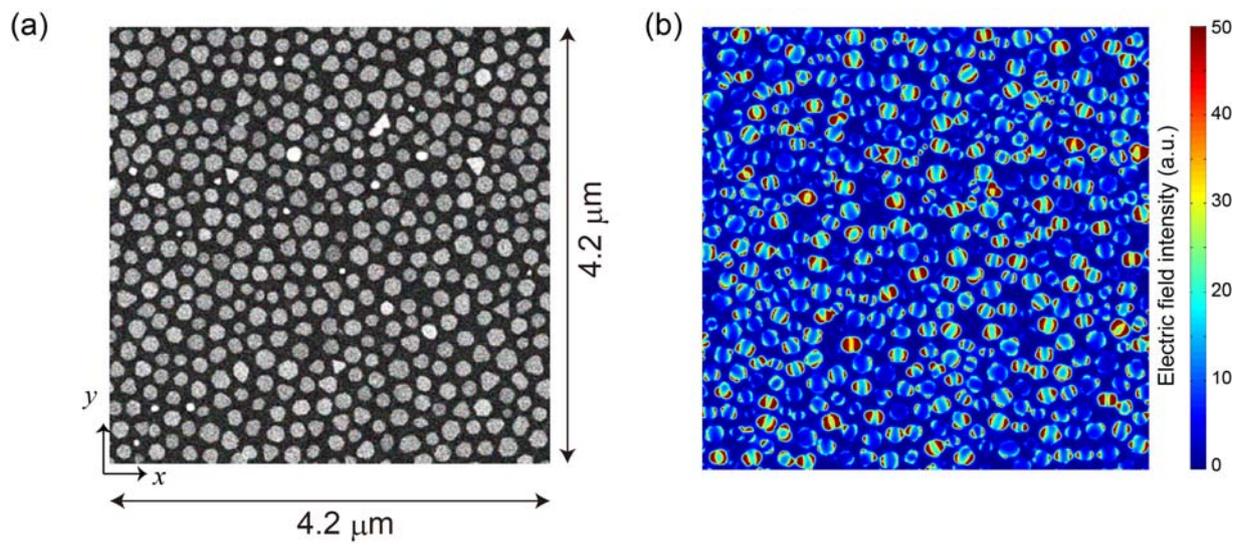

FIG. 1

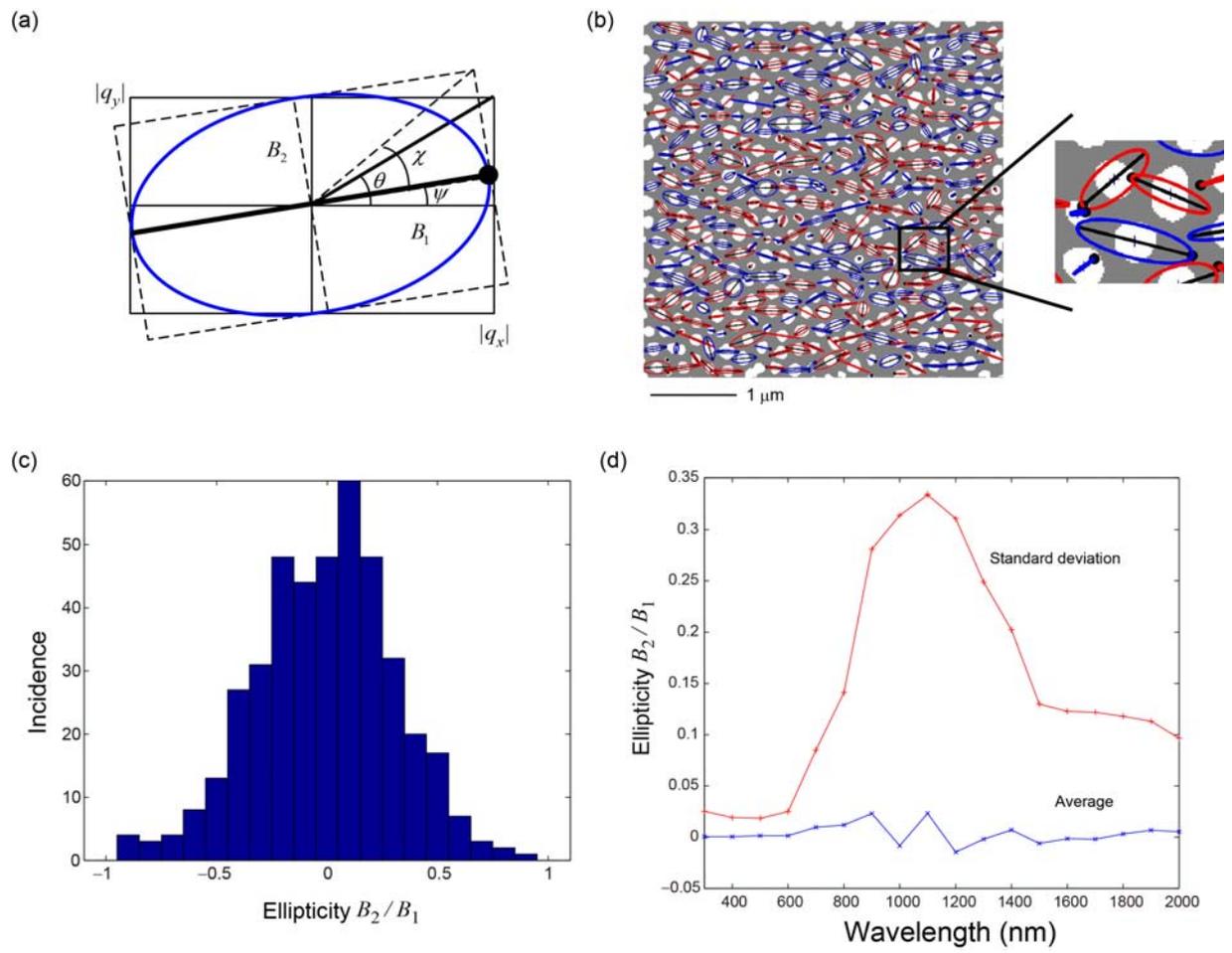

FIG. 2

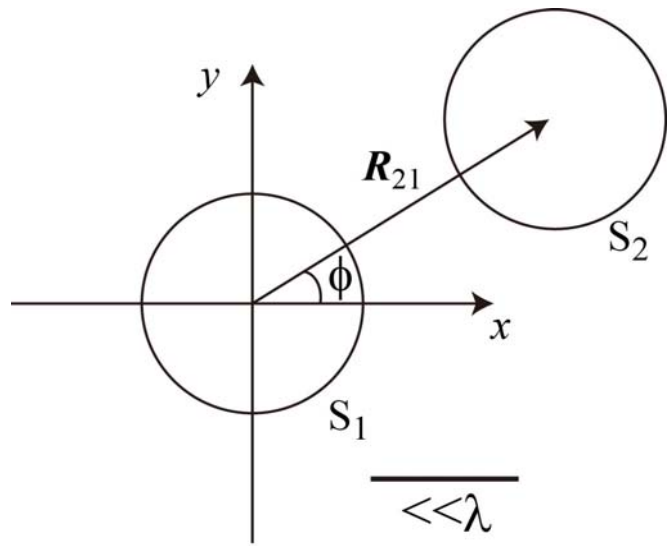

FIG. 3

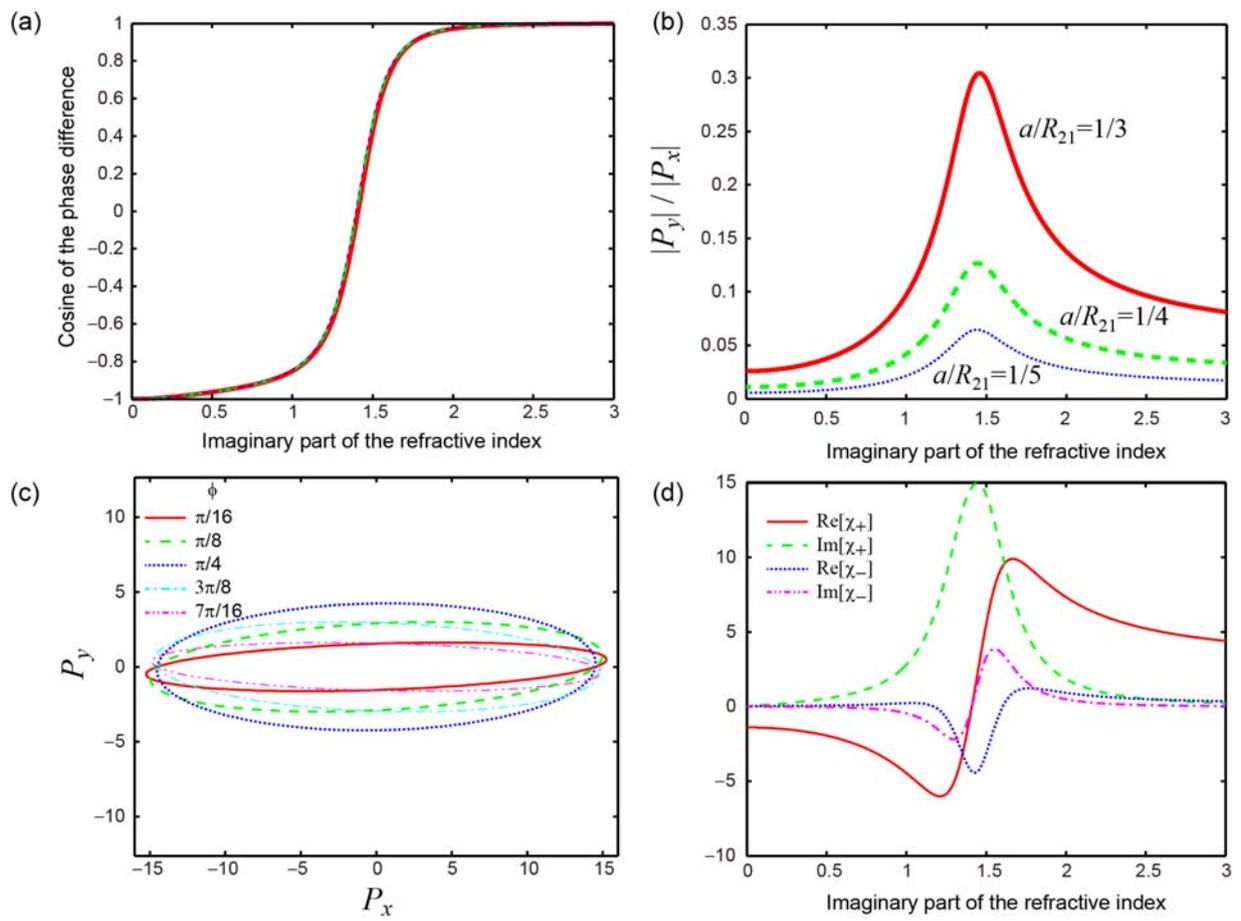

FIG. 4

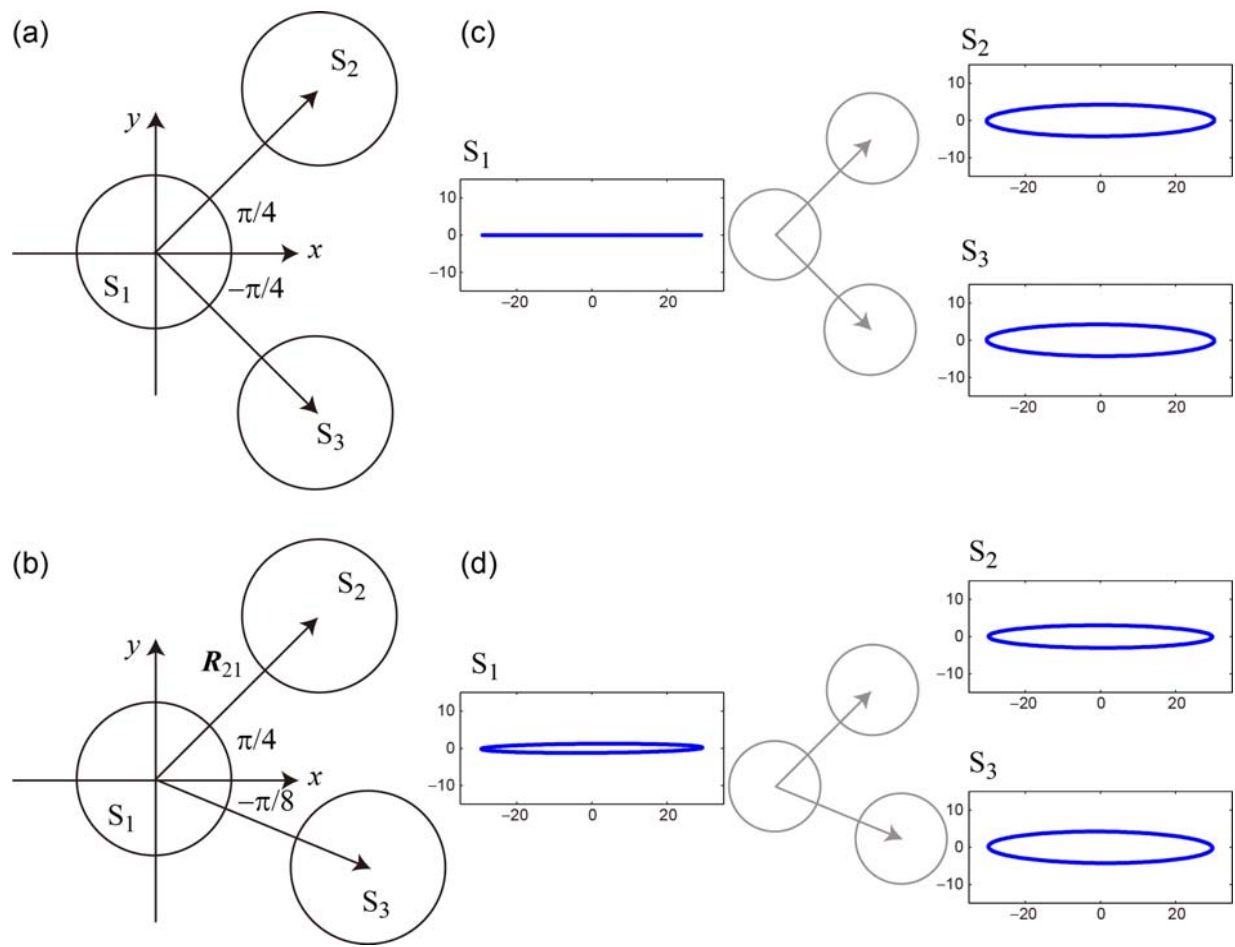

FIG. 5

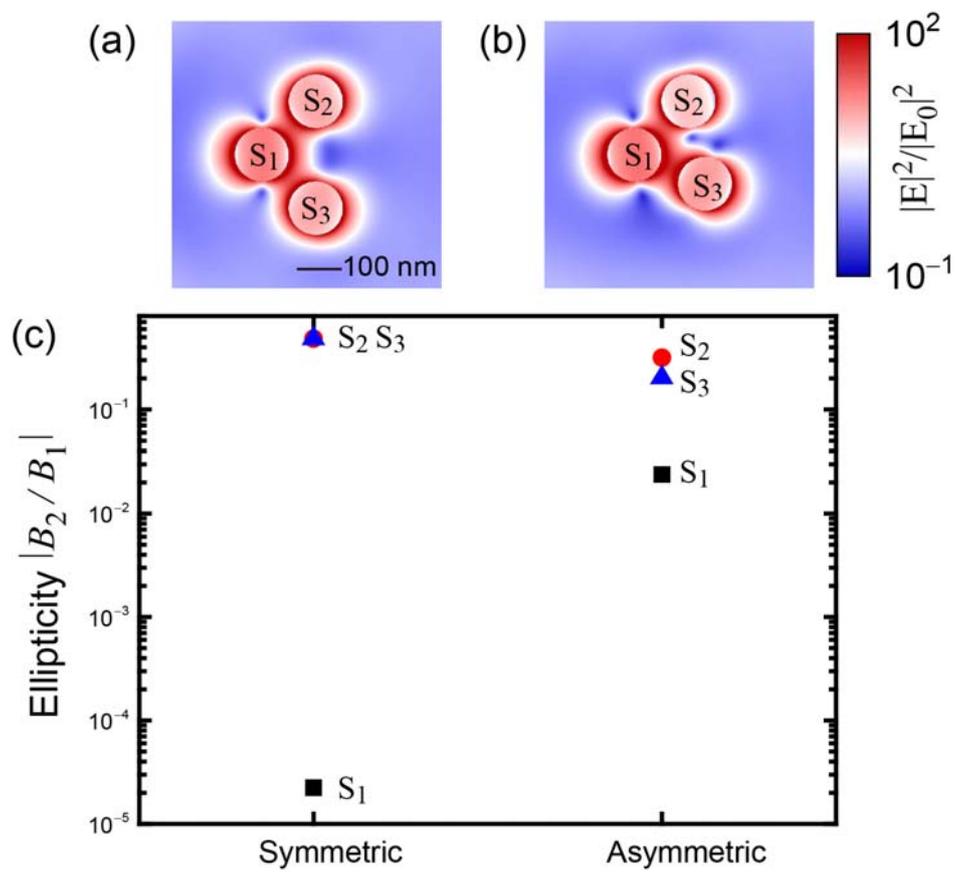

FIG. 6